\documentclass[a4paper,11pt]{article}
\usepackage{pos}

\title{High-precision QCD physics at FCC-ee}

\author*[a,b]{Francesco Giuli}

\affiliation[a]{CERN - European Organization for Nuclear Research,\\
  Espl. des Particules 1, 1211 Meyrin, Switzerland}
\affiliation[b]{On behalf of the FCC collaboration}

\emailAdd{francesco.giuli@cern.ch}

\abstract{The Future Circular Collider (FCC) is a post-LHC project aiming at direct and indirect searches for physics beyond the SM in a new 100 km tunnel at CERN. In addition, the FCC-ee offers unique possibilities for high-precision studies of the strong interaction in the clean environment provided by $e^{+}e^{-}$ collisions, thanks to its broad span of center-of-mass energies ranging from the Z pole to the top-pair threshold, and its huge integrated luminosities yielding $10^{12}$ and $10^8$ jets from Z and W bosons decays, respectively, as well as $10^5$ pure gluon jets from Higgs boson decays. In this contribution, we will summarize studies on the impact the FCC-ee will have on our knowledge of the strong force including: (i) QCD coupling extractions with per-mille uncertainties, (ii) parton radiation and parton-to-hadron fragmentation functions, (iii) jet properties (light-quark-gluon discrimination, $e^{+}e^{-}$ event shapes and multijet rates, jet substructure, etc.), (iv) heavy-quark jets (dead cone effect, charm-bottom separation, gluon $\rightarrow c\bar{c}$, $b\bar{b}$ splitting, etc.); and (v) non-perturbative QCD phenomena (color reconnection, baryon and strangeness production, Bose-Einstein and Fermi-Dirac final-state correlations, etc.).}

\FullConference{%
  41st International Conference on High Energy physics - ICHEP2022\\
  6-13 July, 2022\\
  Bologna, Italy
}

\begin{document}
\maketitle

\section{Introduction}
A crucial aspect for many physics measurements is a precise understanding of Quantum Chromodynamics (QCD). An accurate determination of the strong coupling constant $\alpha_{S}$ is mandatory to improve the precision of the production cross sections and decays calculation. The computation of higher-order N$^{n}$LO and N$^{n}$LL resummation corrections is also central because it can increase the precision in observables predictivity. Another pivotal ingredient is a precise picture of jet substructure, parton showering, hadronization and colour reconnection, whose understanding benefits any hadronic final state.\\
The FCC-ee program~\cite{Agapov:2022bhm}, with its large integrated luminosities and clean environment, offers a rich QCD program. QCD studies with an unprecedented precision can be performed due to the large expected number of events at the FCC-ee of roughly $\sim$ 10$^{11}$ $Z$ at $\sqrt{s}$ = 91 GeV, $\sim$ 10$^{7}$ $W^{+}W^{-}$ at $\sqrt{s}$ = 160 GeV and $\sim$ 10$^{6}$ $ZH$ at $\sqrt{s}$ = 240 GeV.

\section{The strong coupling constant}
The least precisely known of all interaction coupling constant is $\alpha_{S}$, with an overall uncertainty at per-mille level, $\delta\alpha_{S}\sim$ 10$^{-3}$. Currently, $\alpha_{S}$ is determined by comparing 7 experimental observables to perturbative QCD (pQCD) predictions, plus a global average at the $Z$ pole scale. The relevant observable for $e^{+}e^{-}$ collisions are $e^{+}e^{-}$ jet shapes and hadronic $\tau$ leptons and $W/Z$ bosons decays.

\subsection{$\mathbf{\alpha_{S}}$ from $\mathbf{e^{+}e^{-}}$ event shapes and jet rates}  
As already done at LEP~\cite{Dissertori:2009ik}, the thrust ($\tau$) and the $C$-parameter defined in Eq.~\ref{jet_rates} can be used to extract $\alpha_{S}$:
\begin{equation}
\tau = 1 - T = 1 - \max_{\hat{n}}\dfrac{\Sigma|\vec{p}_{i}\cdot\hat{n}|}{\Sigma|\vec{p}_{i}|} \;\;\;\;\;\;\;\;\;\;\;\;\;\;\; C = \dfrac{3}{2}\dfrac{\Sigma_{i,j}|\vec{p}_{i}||\vec{p}_{j}|\sin^2\theta_{i,j}}{(\Sigma_{i}|\vec{p}_{i}|)^{2}},
\end{equation}\label{jet_rates}
with $\theta_{i,j}$ the angle between particle $i$ and $j$ and $\vec{p}_{i,j}$ the momentum respectively. Other quantities which are sensitive to $\alpha_{S}$ are the $n$-jet rates, $R_{n} = \dfrac{\sigma_{n-\mathrm{jet}}}{\sigma_{\mathrm{tot}}}$, and therefore were used to extract the strong coupling constant. The comparison between the experimental measurements and N$^{3}$LO+N$^{2}$LL predictions yields $\alpha_{S}(m_{Z})$ = 0.1171 $\pm$ 0.0027 ($\pm$ 2.6\%). \\
At lower $\sqrt{s}$, the $n$-jet rates up to 7 jets could be studied, while runs at higher $\sqrt{s}$ could be used to study jet rates in regimes where the probability of hard gluon emission increases. Moreover, a better understanding of hadronization mechanism and improvements in logarithmic resummation to N$^{3}$LL for jet rates would allow the extraction of $\alpha_{S}$ at $\delta\alpha_{S}/\alpha_{S} <$ 1\% at the FCC-ee. 

\subsection{$\mathbf{\alpha_{S}}$ from hadronic $\mathbf{\tau}$ decays}
The very precise LEP and B-factories $e^{+}e^{-}\rightarrow\tau^{+}\tau^{-}$ data, together with higher-order pQCD corrections to the hadronic $\tau$ width, allow a remarkably accurate $\alpha_S$ extraction from hadronic $\tau$ decays. The quantity of interest is the ratio of the hadronic $\tau$ width and the electron $\tau$ width, defined as follows:
\begin{equation}
R_{\tau} = \dfrac{\Gamma(\tau^{-}\rightarrow\nu_{\tau}+\mathrm{hadrons})}{\Gamma(\tau^{-}\rightarrow\nu_{\tau}e^{-}\bar{\nu}_{e}} = S_{\mathrm{EW}}N_{\mathrm{C}}\left(1+\Sigma_{n=1}^{4}c_{n}\left(\dfrac{\alpha_{S}}{\pi}\right)^{n}+\mathcal{O}(\alpha_{S}^{5})+\delta_{\mathrm{np}}\right),
\end{equation} 
where $S_{\mathrm{EW}}$ represents the pure electroweak (EW) contribution to the ratio, $N_{\mathrm{C}}$ the number of colours, $c_{n}$ the coefficients of the perturbative expansion, and $\delta_{\mathrm{np}}$ power-suppressed non-perturbative (NP) corrections. Experimentally, this ratio has determined with a $\pm$ 0.23\% precision, and this leads to a determination of $\alpha_{S}(m_{Z}) = 0.1187 \pm 0.0018$ ($\pm$ 1.5\%).\\
The dominant source of theoretical uncertainty in the extraction of $\alpha_{S}$ comes from the discrepancy between the Fixed Order Perturbation Theory (FOPT) and the Contour-Improved Perturbation Theory (CIPT), two different approaches for evaluating the perturbative expansion. Currently, this uncertainty is at the level of $\pm$ 1.5\%. NP correction are also relevant in the determination of $\alpha_{S}$ from hadronic $\tau$ decays. These can be sizeable for $\mathcal{O}(\Lambda_{\mathrm{QCD}}^{2}/m_{\tau}^{2})$ and they can be controlled by new high-precision measurements of the hadronic $\tau$ spectral function.\\
Statistical uncertainty will be negligible at the FCC-ee, considering the $\sim$ 10$^{11}$ $\tau$ produced at the $Z$-pole, and parametric and systematic uncertainties will dominate. To fully exploit this huge statistics, a reduction in the spread of theoretical determinations of $R_{\tau}$ is mandatory. This necessarily implies a better understanding of the discrepancies arising from the CIPT and FOPT comparison. Furthermore, a better determination of the spectral functions entering the $R_{\tau}$ calculation is compulsory, and this can be achieved exploiting new data coming from Belle II or the FCC-ee itself. In this way, the uncertainty on $\alpha_{S}$ can be reduced well below the current $\delta\alpha_{S}/\alpha_{S}\sim$ 1\% level.

\subsection{$\mathbf{\alpha_{S}}$ from hadronic $\mathbf{W}$ boson decays}\label{sec:Wwidth}
\begin{figure}[t!]
\begin{center}
\includegraphics[width=0.95\textwidth]{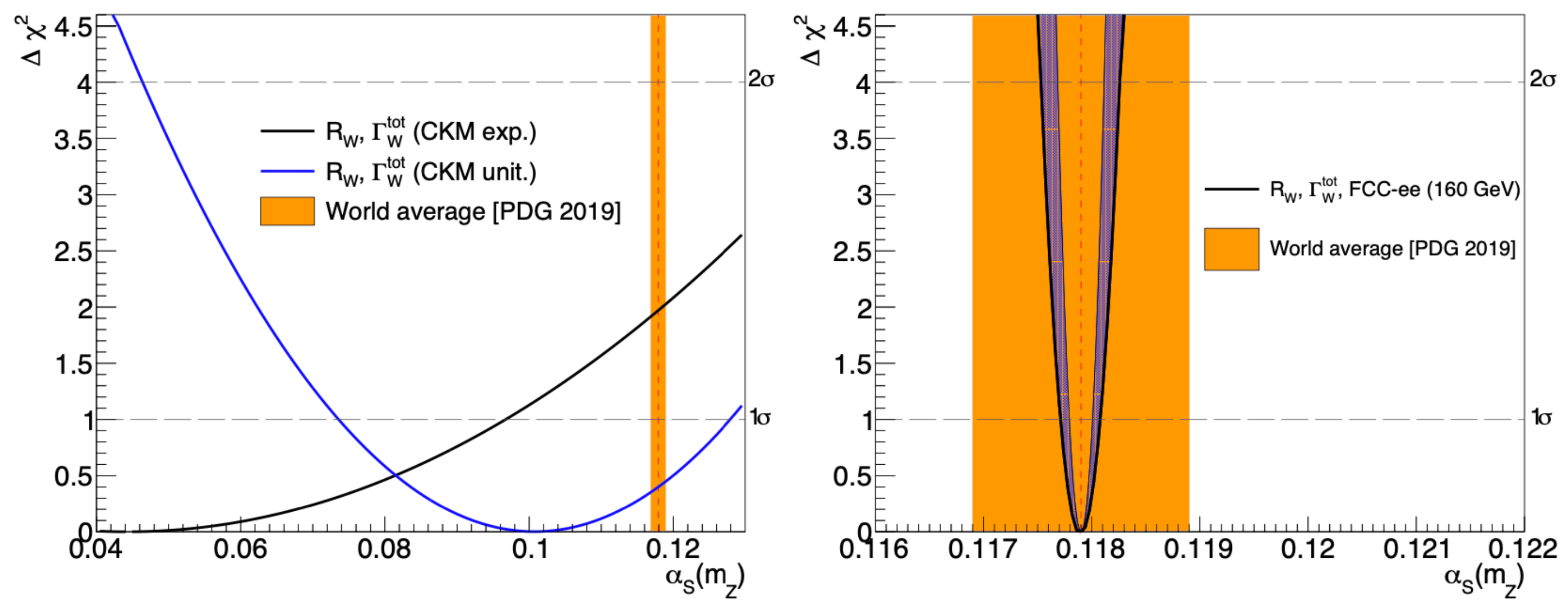}
\end{center}
\caption{$\Delta\chi^{2}$ fit profiles of the $\alpha_{S}(m_{Z})$ extracted from the combined N$^{3}$LO analysis of the total $W$ width ($\Gamma_{W}^{\mathrm{tot.}}$) and hadronic-to-leptonic $W$ decay ratio ($R_{W}$), compared to the current $\alpha_{S}(m_{Z})$ world average (vertical orange band). Left: Extraction with the present W data assuming (blue curve) or not (black curve) CKM unitarity. Right: Extraction expected at the FCC-ee, with the total (experimental, parametric, and theoretical in quadrature) uncertainties (outer parabola) and with the experimental uncertainties alone (inner parabola). These plots are taken from Ref.~\cite{dEnterria:2020cpv}.} 
\label{fig:Wwidth}
\end{figure}
\vspace{-1mm}
Analogously to the case of the hadronic $\tau$ decays, the extraction of $\alpha_{S}$ from hadronic $W$ boson decays can be performed considering the ratio of the hadronic width to the lepton with, as described in Eq.~\ref{Whad}
\begin{equation}
R_{W}(Q)=\dfrac{\Gamma_{W}^{\mathrm{had.}}(Q)}{\Gamma_{W}^{\mathrm{lep.}}(Q)}=R_{W}^{\mathrm{EW}}\left(1+\Sigma_{i=1}^{4}a_{i}(Q)\left(\dfrac{\alpha_{S}(Q)}{\pi}\right)^{i}+\mathcal{O}(\alpha_{S}^5)+\delta_{\mathrm{mix}}+\delta_{\mathrm{np}}\right)
\end{equation}\label{Whad}
with $R_{W}^{\mathrm{EW}}$ representing the pure EW contribution to the ratio, $a_{i}(Q)$ the coefficients of the perturbative expansion, $\delta_{\mathrm{mix}}$ the mixed QCD+EW corrections, and $\delta_{\mathrm{np}}$ the power-suppressed NP corrections. 
$\alpha_{S}$ is then extracted at N$^{3}$LO from a simultaneous fit of 2 $W$ boson pseudo-observables~\cite{dEnterria:2020cpv}: $R_{W}$ and $\Gamma_{W}^{\mathrm{tot.}}$. With the assumption of CKM unitarity, a value of $\alpha_{S}(m_{Z}) =$ 0.101 $\pm$ 0.027 is obtained (with negligible theoretical and parametric uncertainties), as depicted in Fig.~\ref{fig:Wwidth} (left). The large uncertainty is mostly due to the poor experimental knowledge of $R_{W}$ and $\Gamma_{W}^{\mathrm{tot.}}$, which have been measured in $e^{+}e^{-}\rightarrow W^{+}W^{-}$ LEP events. If CKM unitarity is not assumed, the resulting value of the strong coupling constant is basically unconstrained, as shown in Fig.~\ref{fig:Wwidth} (left).\\
At the FCC-ee, the uncertainties on $R_{W}$ and $\Gamma_{W}^{\mathrm{tot.}}$ will be largely reduced, thanks to the high statistics at the $WW$ threshold. With a factor of 10 reduction of the theoretical uncertainties due to missing $\alpha_{S}^{5}, \alpha^{3}, \alpha\alpha_{S}^{2}$ and $\alpha^{2}\alpha_{S}$ corrections, a final QCD coupling extraction of $\alpha_{S}(m_{Z}) =$ 0.11790 $\pm$ 0.00023 with 2 per-mille total error is possible, as illustrated in Fig.~\ref{fig:Wwidth} (right).

\subsection{$\mathbf{\alpha_{S}}$ from hadronic $\mathbf{Z}$ boson decays}
\begin{figure}[t!]
\begin{center}
\includegraphics[width=0.95\textwidth]{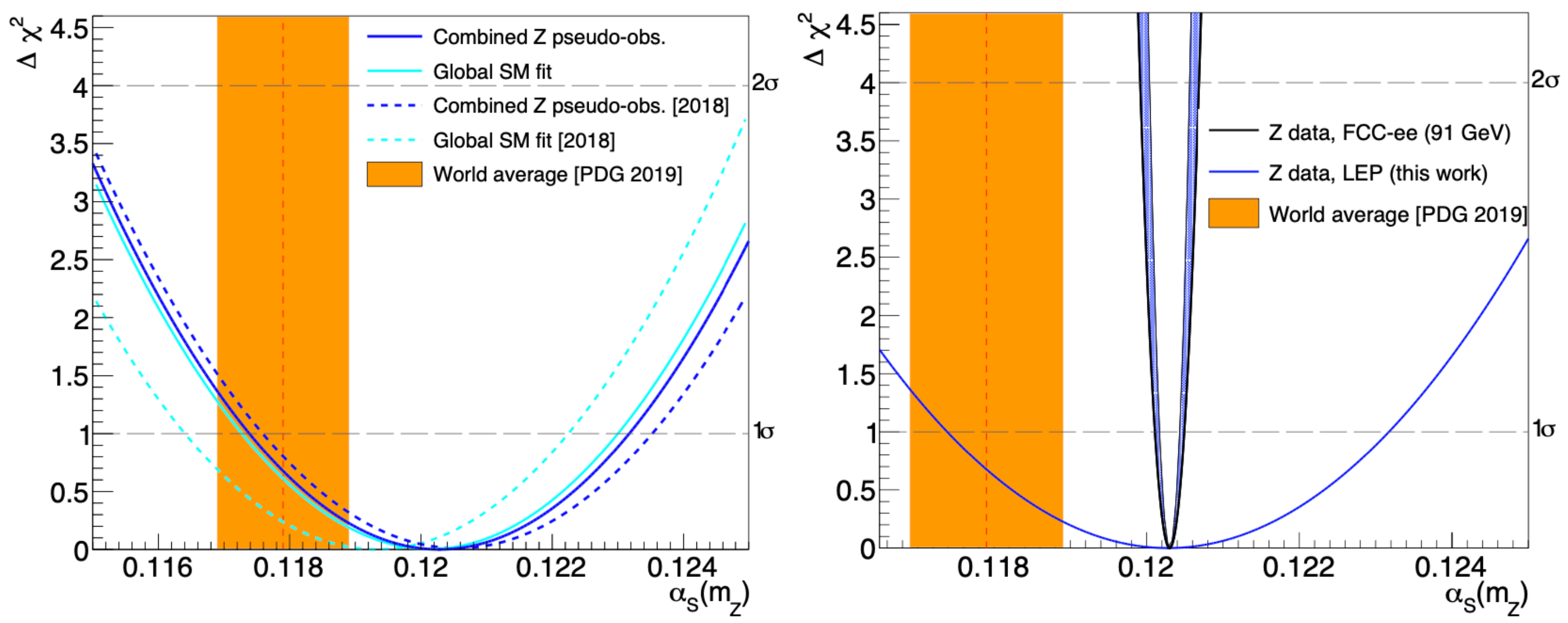}
\end{center}
\caption{$\Delta\chi^{2}$ fit profiles of $\alpha_{S}(m_{Z})$ extracted from the combined $Z$ pseudo-observables analysis and/or the global SM fit compared to the current world average (orange band). Left: Current results (solid lines) compared to the previous 2018 fit (dashed lines). Right: Extraction expected at the FCC-ee - with central value (arbitrarily) set to $\alpha_{S}(m_{Z})$ = 0.12030 and total (experimental, parametric, and theoretical in quadrature) uncertainties (outer parabola) and experimental uncertainties alone (inner parabola) – compared to the present one from the combined $Z$ data (blue line). These plots are taken from Ref.~\cite{dEnterria:2020cpv}.} 
\label{fig:Zwidth}
\end{figure}
Following the same procedure described in Sec.~\ref{sec:Wwidth}, $\alpha_{S}$ can be extracted at N$^{3}$LO from a simultaneous fit of 3 $Z$ boson pseudo-observables~\cite{dEnterria:2020cpv}: $R_{Z}$, $\Gamma_{Z}^{\mathrm{tot.}}$ and $\sigma_{Z}^{\mathrm{had.}}$, yielding $\alpha_{S} =$ 0.1203 $\pm$ 0.0029 ($\pm$ 2.3\%), as depicted in Fig.~\ref{fig:Zwidth} (left).\\
Having 10$^{5}$ times more $Z$ bosons than at LEP, together with an exquisite systematic and parametric precision would allow a remarkable improvement in the theoretical predictions of the $Z$ boson pseudo observables, and therefore a reduction in the strong coupling uncertainty by almost 2 orders of magnitude. This experimental precision has to be matched by a reduction in the theoretical uncertainties by almost a factor of 5 by computing missing $\alpha_{S}^{5}, \alpha^{3}, \alpha\alpha_{S}^{2}$ and $\alpha^{2}\alpha_{S}$ corrections. In this way, $\alpha_{S}$ can be extracted with a 2 per-mille accuracy, namely $\alpha_{S}(m_{Z})$ = 0.11790 $\pm$ 0.00023, as reported in Fig.~\ref{fig:Zwidth} (right).

\section{Jet substructure}
\begin{figure}[t!]
\begin{center}
\includegraphics[width=0.95\textwidth]{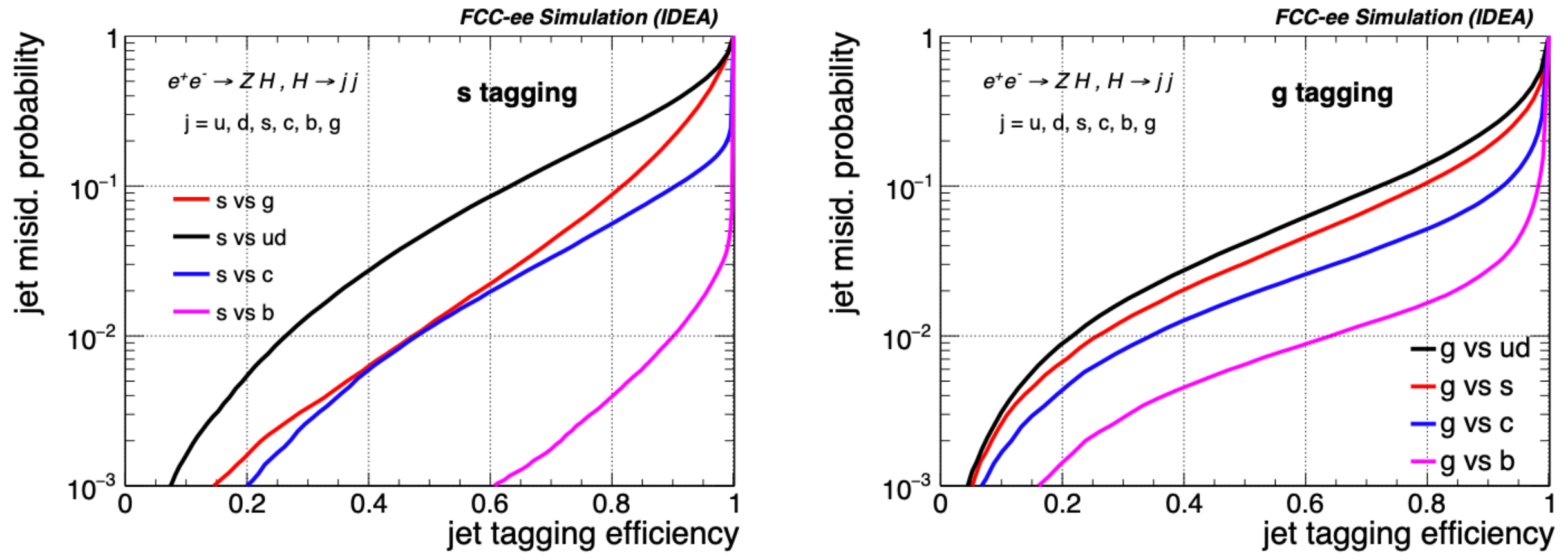}
\end{center}
\caption{Evaluation of \texttt{ParticleNetIdea} performance in terms of a receiver operating characteristic (ROC) curve for the identification of different jet flavours i.e., $s$ (left), and $g$ (right). The different jet flavours considered background are indicated on the labels. The IDEA detector configuration is used. These plots are taken from Ref.~\cite{Bedeschi:2022rnj}.} 
\label{fig:tagging}
\end{figure}
Jet substructure studies play a crucial role in improving our knowledge of parton shower (PS) and hadronization mechanism. In particular, jet angularities~\cite{Larkoski:2014pca}, defined as $\lambda_{\beta}^{\kappa}=\Sigma_{i\in\mathrm{jet}}z_{i}^{\kappa}\theta_{i}^{\beta}$ (with $z_{i}$ and $\theta_{i}$ representing the energy fraction and angular distance to jet axis of constituent $i$), constitute an intriguing starting point. The parameters $\kappa\geq$ 0 and $\beta\geq$ 0 regulate the energy and angular weighting respectively. Multiplicity ($\kappa$ = 0, $\beta$ = 0), width ($\kappa$ = 1, $\beta$ = 1), mass ($\kappa$ = 1, $\beta$ = 2), $p_{\mathrm{T}}^{\mathrm{D}}$ ($\kappa$ = 0, $\beta$ = 2) and Les Houches Angularity ($\kappa$ = 1, $\beta$ = 0.5) are the most common examples. Specifically, this last quantity offers an incredible opportunity to study different PS algorithms between generators.\\
The FCC-ee would be crucial in addressing such differences in PS and hadronization modelling. For example, the gluon radiation patters could be studied exploiting the expected 10$^6$ $e^{+}e^{-}\rightarrow ZH(\rightarrow gg)$ events, together with the $e^{+}e^{-}\rightarrow Z\rightarrow b\bar{b}g$ events (assuming that $b$-jets are tagged with high efficiency. Therefore, these studies conducted at the FCC-ee would lead directly to improved MC tuning, together with a better understanding of NP QCD. 

\section{Quark-gluon tagging}
One of the most exciting (but challenging) prospects in $pp$ collisions is light-quark gluon discrimination. Being able to efficiently identify the flavour of the parton which initiates the jet is critical for the success of the physics program of future EW factories~\cite{Azzi:2021gwg}. An accurate light quark-gluon discrimination would allow precise Beyond the Standard Model (BSM) searches for signals without leptons, $b$- or top-quarks, as well as would produce an enhancement of light quark-rich signals i.e. $t\bar{t}H$ or pure EW $W/Z$ + jets.\\
Recently, a new generation of advanced machine learning based jet tagging algorithms has been developed~\cite{ATLAS:2015thz,CMS:2020poo,ATLAS:2017pe,Bols:2020bkb}, bringing almost 2 orders of magnitude improvement in background rejection when comparing to the traditional approaches in Heavy Flavour and gluon tagging. In particular, within the context of the FCC-ee, the \texttt{ParticleNetIdea}~\cite{Bedeschi:2022rnj} has been developed, and Figure~\ref{fig:tagging} shows its high performances in discriminating light quark jets from $s$-quark (left) and gluons (left).

\section{Conclusion}
To fully exploit present and future collider programs, a precise understanding of both perturbative and NP QCD is highly needed. At the FCC-ee, a plethora of unique QCD studies would be possible. Among them, the most relevant are the extraction of the strong coupling constant $\alpha_{S}$ from jet event shapes and hadronic $\tau/W/Z$ decays with a per mille level accuracy and jet substructure studies, which could greatly improve our current knowledge of parton shower and hadronization. Thanks to the large pure quark/gluon samples in the extremely clean environment of a lepton collider, precise quark-gluon discrimination studies would be carried out with a much better discriminating power than the one in $p\bar{p}/pp$ collisions. Finally, due to the large number of expected $e^{+}e^{-}\rightarrow W^{+}W^{-}$, the huge statistics ($\times$ 10$^{4}$ LEP) could be exploited to measure the $W$ boson mass, $m_{W}$, both (semi-)leptonically and hadronically to constrain colour reconnection at the 1\% level or below.

\end{document}